\pdfoutput=1 
\documentstyle[12pt,graphicx,natbib,hyperref]{article}
\def\sun{\hbox{$\odot$}}

\def\sun{\ast{\odot}}
\def\massbolo{M^\odot_{K_S}}
\def\apj{ApJ\,  }
\def\apjs{ApJS  }
\def\cjaa{Chinese J. Astron. Astrophys.  }

\def\mnras{MNRAS\,  }
\def\pasa{PASA  }

\begin{document}
\title{
The intergalactic Newtonian gravitational 
field and the shell theorem 
}

\author
{L. Zaninetti             \\
Dipartimento di Fisica , \\
           Via Pietro Giuria 1   \\
           10125 Torino, Italy
}

\maketitle
\section*{}
The release  of the 2MASS Redshift Survey (2MRS),
with it's 
44599  galaxies 
allows the deduction of the galaxy's masses 
in nearly complete sample.
A cubic box with  side of 37 Mpc
containing 2429 galaxies is  extracted  and  
the Newtonian gravitational field  
is  evaluated   both at the center
of the box 
as well   in  101 x 101 x 101   grid points of the box.
The obtained results  are then discussed at the 
light of the shell theorem 
which states that at the internal  of a sphere 
the gravitational field is zero.
\\
keywords                    \\
{
methods: statistical;
cosmology: observations;
(cosmology): large-scale structure of the Universe
}

\section{Introduction}

The determination of  the gravitational  field 
in cosmology   oscillates between the Newton law
and various type of modifications to this law.
The reference formula is the Newtonian 
force 
\begin{equation}
F = - G \frac {m M}{r^2} 
\quad ,
\end{equation}
where $G$ is the gravitational constant, $M$ the first mass, 
$m$ the second mass
and $r$ the distance between the two masses.
The enormous  progresses in  the observations
of the spatial distribution  of galaxies 
point  toward  a cellular structure, i.e. the 
galaxies  are situated on the surfaces of bubbles,
rather than to be aggregated in a random structures,
see  \cite{Coil2012}.
In the limiting case in which all the galaxies 
were  situated on the surfaces  of spheres the 
gravitational forces should be zero due to the shell 
theorem or nearly zero due to the fact that the 
galaxies  are distributed in a discrete way 
rather than in a continuous  way.
This paper  describes 
in Section \ref {observations}
two astronomical catalogs  which allow 
to calibrate the size of the cosmic voids. 
Section \ref{secphotometric}
is devoted to the study  of  the
photometric properties of 
of a nearly spherical  distribution  of galaxies
and to a careful  analysis 
of completeness connected with  the selected 
astronomical  catalog.
Section \ref{sectionshell} contains the evaluation 
of the Newtonian gravitational field in a box 
of 37 $Mpc$ in which the boundary 
conditions are properly evaluated.
Section  \ref{secvoronoi}  reports  a comparison 
between  three ideal  structures   
and   a real void as extracted from  a slices oriented  
catalog.

\section{Observations}
\label{observations} 

This section processes  the 
Sloan Digital Sky Survey Data Release 7 (SDSS DR7),
see  \cite{Abazajian2009}, 
and  the 
2MASS Redshift Survey (2MRS), 
see \cite{Huchra2012}.  

\subsection{Observed statistics of the voids}
\label{sec_statistics}

The distribution  of the effective radius  
between the galaxies
of  SDSS DR7
has been reported  in 
\cite{Vogeley2011}.
This catalog  contains   1054  voids  
and Table \ref{statvoids} reports their basic
statistical  parameters. 
\begin{table}
 \caption[]
{
The statistical  parameters 
of the effective radius in  SDSS DR7.
}
 \label{statvoids}
 \[
 \begin{array}{lc}
 \hline
 \hline
 \noalign{\smallskip}
parameter                   &   value                          \\ \noalign{\smallskip}
mean                        &  18.23h^{-1}~ Mpc   \\ \noalign{\smallskip}
variance                    &  23.32h^{-2}~ Mpc^2 \\ \noalign{\smallskip}
standard~ deviation         &  4.82h^{-1} ~ Mpc   \\ \noalign{\smallskip}
kurtosis                    &  0.038        \\ \noalign{\smallskip}
skewness                    &  0.51         \\ \noalign{\smallskip}
maximum ~value              &  34.12h^{-1}~ Mpc   \\ \noalign{\smallskip}
minimum ~value              &  9.9h^{-1}~   Mpc   \\ \noalign{\smallskip} \hline
 \hline
 \end{array}
 \]
 \end {table}

\subsection{The 2MASS}

\label{catalogue}
\begin{figure}
\begin{center}
\includegraphics[width=6cm]{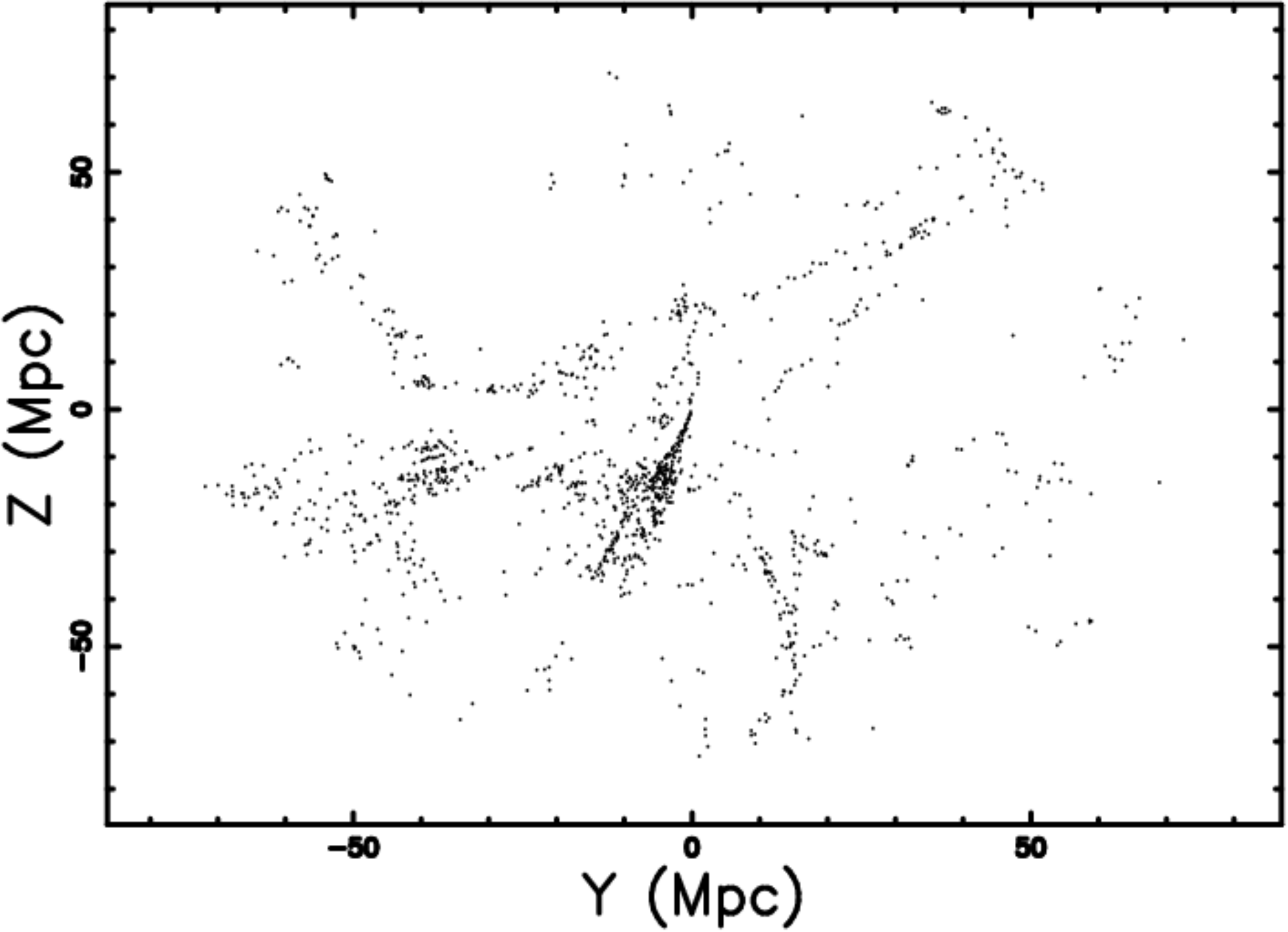}
\end {center}
\caption
{
Cut  of the 3D spatial distribution  of 2MRS  
in the $X=0$ plane  when  $\Delta =$ 10 Mpc,
the squared box has a side of  148 $Mpc$  
and  $N_G$ = 1244.
}
          \label{2mrs_cut}%
    \end{figure}

The 2MASS   is a catalog
of galaxies which has instruments in the near-infrared
J, H and K-bands
(1-2.2 $\mu$\ m)
 and therefore detects  the galaxies
in the so called ``Zone of Avoidance,''
see \cite{Jarrett2004,Huchra2007}.
At  the moment  of writing
the  2MRS
consists of  44599  galaxies
with redshift in the interval $0\leq z\leq 0.09$,
see  \cite{Huchra2012}.
The catalog  contains the galactic  latitude,
the galactic longitude and the 
expansion velocity; from these 
three parameters is possible  
to deduce the Cartesian coordinates, $X,Y$ and $Z$ 
expressed in $Mpc$.
Figure \ref{2mrs_cut} reports a  cut
of  a given thickness 
of 2MRS  where $\Delta$ express the thickness of the cut 
and $N_G$ the number of selected galaxies.

%
\section{Photometric  properties}

\label{secphotometric}

\begin{figure}
\begin{center}
\includegraphics[width=6cm]{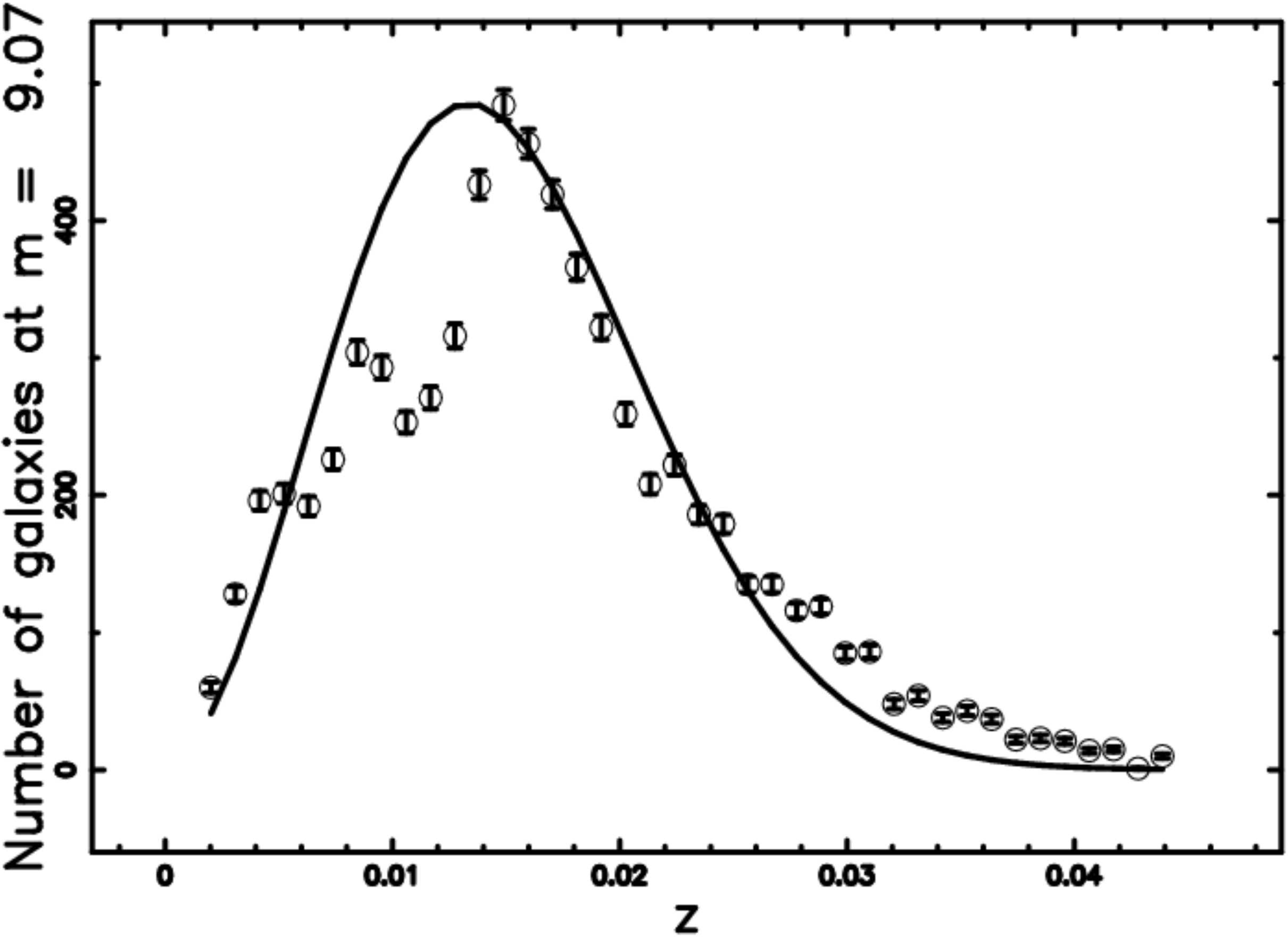}
\end {center}
\caption{
The galaxies  of the 2MRS with
$ 8.48  \leq  m  \leq 10.44  $  or
$  1202409  \frac {L_{\sun}}{Mpc^2} \leq
f \leq  7267112 \frac {L_{\sun}}{Mpc^2}$
are  organized in frequencies versus
heliocentric  redshift,
(empty circles);
the error bar is given by the square root of the frequency.
The maximum frequency of observed galaxies is
at  $z=0.015$.
The full line is the theoretical curve  
generated by  $\frac{dN}{d\Omega dz df}(z;z_{crit},c,H_0)$.
In this plot $\massbolo$ = 3.39,
$h$ = 0.7,
$M^*$=-24.87,
$\alpha$ =-0.98,
$\Phi^*$=0.0037,
$\chi^2=721$ and the number of bins 40.
}
          \label{maximum_flux}%
    \end{figure}

This section reviews  the photometric maximum 
in the framework 
of the luminosity  function for galaxies  and 
the Malmquist  bias 
which fixes the concept  of complete  sample.
A model for the luminosity of galaxies
is  the Schechter function,
$\Phi (L;L^*, \alpha, \Phi )$, 
where $\alpha$ sets the slope 
for low values of $L$, $L^*$ is the
characteristic luminosity, 
and $\Phi^*$ is a normalization,
see 
\citep[] [Eqn. (55)] {Zaninetti2010g}.
This
function  was suggested  by \cite{schechter} and
the distribution in absolute magnitude 
,$\Phi (M;M^*, \alpha, \Phi )$, 
can be found
in  \citep[] [Eqn. (56)] {Zaninetti2010g}
where $M^*$ is the characteristic magnitude
as derived from the
data.
The parameters  of the Schechter function
concerning the 2MRS as well the bolometric 
luminosity, $\massbolo $, 
can be found in \cite{Cole2001} and
are reported in   Table~\ref{parameters}.
\begin{table}
 \caption[]
{
The parameters of the Schechter function
and bolometric magnitude
for the 2MRS   in the $K_s-band$.
}
 \label{parameters}
 \[
 \begin{array}{lc}
 \hline
 \hline
 \noalign{\smallskip}
parameter            & 2MRS                            \\ \noalign{\smallskip}
M^* - 5\log_{10}h ~ [mags] &  ( -23.44 \pm 0.03) \\ \noalign{\smallskip}
\alpha                     &   -0.96  \pm 0.05                 \\ \noalign{\smallskip}
\Phi^* ~[h^3~Mpc^{-3}]     &   ((1.08   \pm 0.16)10^{-2})    \\ \noalign{\smallskip}
\massbolo                  &   3.39                            \\ \noalign{\smallskip}
h                          &   0.7                              \\ \noalign{\smallskip}
 \hline
 \hline
 \end{array}
 \]
 \end {table}
The number of galaxies at a given flux  $f$
as  a function of the redshift  $z$,
$\frac{dN}{d\Omega dz df}(z;z_{crit},c,H_0)$, 
can be found in 
\citep[] [Eqn.(1.104)] {pad}
or  
\citep[] [Eqn. (6)] {Zaninetti2010g},
where $d\Omega$, $dz$, and  $ df $
are the differentials of
the solid angle, the red-shift, and the flux, respectively,
$z_{crit}$ a  parameter,
$H_0$ the  Hubble constant 
and  $c$ is the velocity of light.
The number of galaxies at a given flux
has a maximum  at  $z=z_{max}(z_{crit}, \alpha) $, 
see \citep[] [Eqn. (8)] {Zaninetti2010g}.
Figure~\ref{maximum_flux}
reports the number of  observed  galaxies
in  the 2MRS  catalog at a given
apparent magnitude  and
the theoretical curve  as represented by
$\frac{dN}{d\Omega dz df}(z;z_{crit},c,H_0)$.
The merit function $\chi^2$
can be computed as
\begin{equation}
\chi^2 =
\sum_{j=1}^n ( \frac {n_{theo}(z) -
                      n_{astr}(z) } {\sigma_{n_{astr}(z)}})^2
\quad ,
\label{chisquare}
\end{equation}
where   $n$ is number of data, 
the two 
indexes $theo$ and $astr$
stand for theoretical and astronomical, respectively and
 ${\sigma_{n_{astr}(z)}}^2$  is the variance of
 the astronomical number of data; the
obtained value is  reported in the caption of Figure
\ref{maximum_flux}.

\begin{figure}
\begin{center}
\includegraphics[width=6cm]{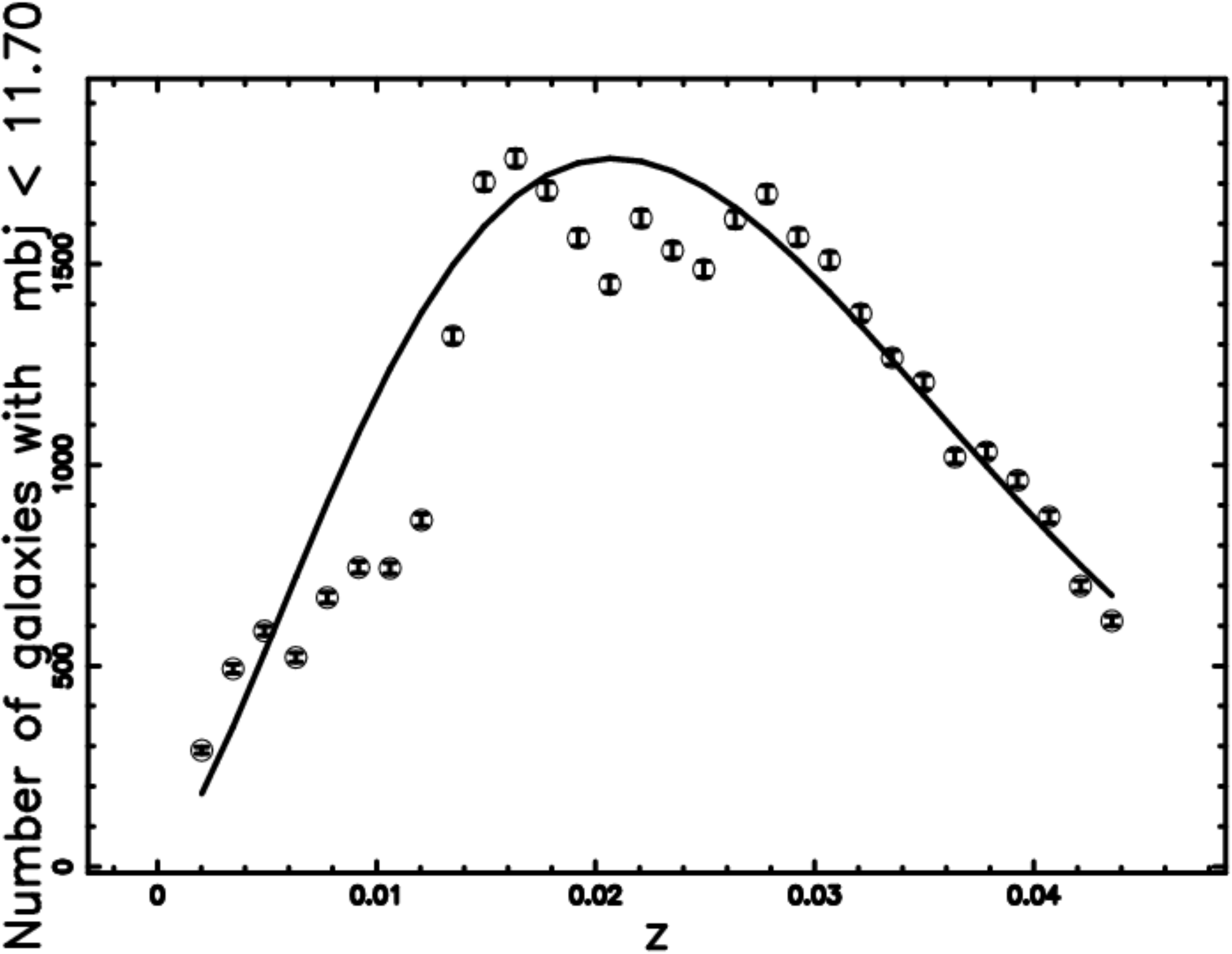}
\end {center}
\caption{
All the  galaxies  of the 2MRS 
with $m_{ks} < $ 11.75 
are  organized in frequencies versus
heliocentric  redshift,
(empty circles);
the error bar is given by the square 
root of the frequency.
The maximum frequency of all observed galaxies is
at  $z=0.017$.
The full line is the theoretical curve  
generated by 
$\frac{dN}{d\Omega dz df}(z;z_{crit},c,H_0)$.
In this plot $\massbolo$ = 3.39, $h$ = 0.7,
$M^*$=-23.97,
$\alpha$ =-0.96,
$\Phi^*$=0.0037,
$\chi^2=1267$ and the number of bins is 30.
}
          \label{maximum_flux_tutte}%
    \end{figure}

The total number of galaxies in the 2MRS
as function of $z$ 
is reported in Figure~\ref{maximum_flux_tutte}  
as well 
as the theoretical curve   represented 
by the numerical integration of 
$\frac{dN}{d\Omega dz df}(z;z_{crit},c,H_0)$.

\begin{figure*}
\begin{center}
\includegraphics[width=6cm]{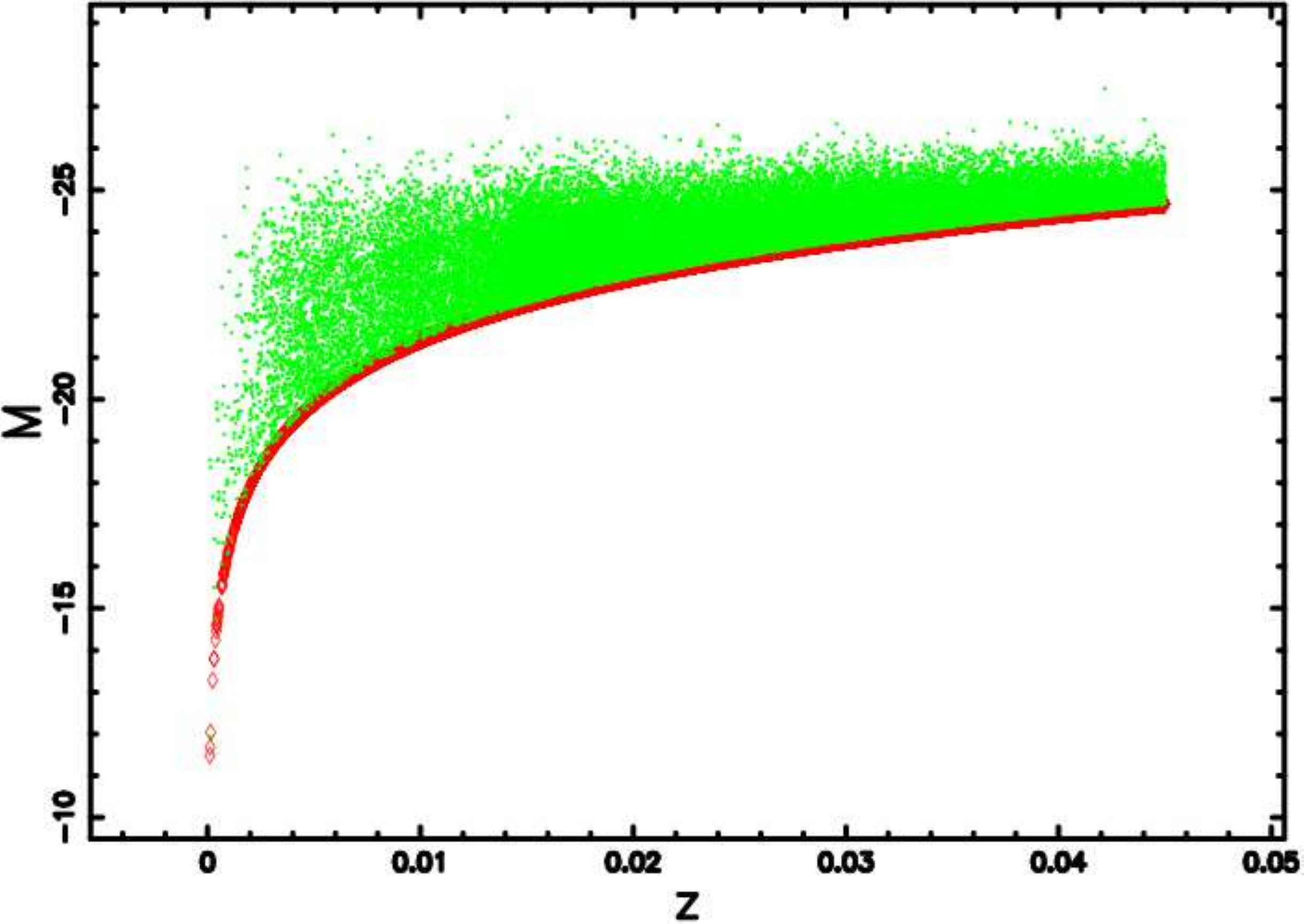}
\end {center}
\caption{
The absolute magnitude $M$ of
36464 galaxies belonging to the 2MRS
when $\massbolo$ = 3.39 and
$H_{0}=70 \mathrm{\ km\ s}^{-1}\mathrm{\ Mpc}^{-1}$
(green points).
The upper theoretical curve as represented by
$M_L(m_L;z,H_0)$ is reported as the
red thick line when $m_L$=11.75.
}
 \label{bias}%
 \end{figure*}

The mass of a galaxy can be evaluated once the mass luminosity
ratio, $R$ is given by
\begin{equation}
R = \langle   \frac{M}{L} \rangle
\label{ratior}
\quad .
\end{equation}
Some values of  $R$ are now reported:
$ R \leq 20$ by~\cite{kiang1961} and \cite{Persic_1992},
$R =20$ by~\cite{pad} and  
$R =5.93 $ by~\cite{vandermarel1991}.
The
 Malmquist bias,
see \cite{Malmquist_1920,Malmquist_1922},
 was originally applied
to the stars and later on 
 to the galaxies by \cite{Behr1951}.
The observable absolute magnitude,
$M_L(m_L;z,H_0)$, 
as a function of the
limiting apparent magnitude, $m_L$, 
can be found  in  
\citep[] [Eqn. (51)] {Zaninetti2010g}.
The bias  predicts, from a theoretical
point of view, an upper limit for  the maximum 
absolute  magnitude which  can be observed in a
catalog of galaxies characterized by a given limiting
magnitude and Figure~\ref{bias} reports such a curve
as well  the galaxies of the 2MRS.

The limiting magnitude of the 2MRS is
$m_L$=11.75  and  therefore 
the 2MRS  is complete
for $z\leq0.0025$.
For values  of $z$ greater than 
this value  the observed sample is not complete
and  we can introduce the  efficiency,
 $\epsilon_s(z;M_{max},M_{min},m_L,c,h)$,
where $M_{max}$ and $M_{min}$ are the
maximum and minimum
absolute
magnitude of  the considered catalog  and 
$h=H_0/100$, see \citep[] [Eqns. (51-53)] {Zaninetti2010g}.
As an example  when  $z\approx 0.017$ the sample  
covers the 52$\%$ of the range in  absolute magnitude.

\section{The gravitational field}

\label{sectionshell}
We now explore a connection with the shell theorem.
Let us consider a spherically  symmetric surface 
of radius    $a$ on which  
a total mass $M$ is disposed in a uniform way.
The mass per unit area $\mu$ is
\begin{equation}
\mu =  \frac{M}{4\pi a^2}
\quad .
\end{equation}  
The force at the internal of 
the spherical  surface
is 
\begin{equation}
\Gamma =0  \quad  r <  a  
\quad ,
\end {equation}
see  Eq. (11.24) in \cite{Alonso1992}
and  therefore 
the shell theorem can be formulated 
"A uniform shell of matter 
exerts no gravitational force  on a particle situated 
inside a shell".
We now compute   
the force  on the center 
($x$=0,$y$=0,$z$=0)
of a hemisphere  which resides on the positive  z-axis.
The  vectorial intensity  of the  field  is 
\begin{equation}
d {\bf \Gamma}  
=G \frac{dm}{a^2}
\quad  ,
\end{equation}  
being 
\begin{equation}
dm = \mu  a^2  d\Omega
\quad ,
\end{equation}
where  
$d \Omega = \sin \theta d \theta d \phi$ 
is the  solid angle  with 
$0 \leq \phi \leq 2 \pi$ and    
$0 \leq \theta \leq \frac {\pi}{2}$.
The three Cartesian components  
of the field  in the 3D case  are 
\begin{eqnarray}
\Gamma_z = \int_0^{2\pi} d \phi 
           \,\int_{\frac{\pi}{2}}^0 
G \, \mu \, sin \theta \,cos \theta 
\, 
d \theta 
\nonumber  \\
\Gamma_x = \int_0^{2\pi} d \phi 
           \,\int_{\frac{\pi}{2}}^0 
G \, \mu \, sin \theta \,sin \theta \, \cos \phi
\, 
d \theta 
\\
\Gamma_y = \int_0^{2\pi} d \phi 
           \,\int_{\frac{\pi}{2}}^0 
G \, \mu \, sin \theta \,sin \theta \, \sin \phi
\, 
d \theta 
\nonumber 
\quad . 
\end {eqnarray}
The integration gives  in the 3D case
for the  three forces at the center  
\begin{eqnarray}
\Gamma_z = \frac{G M} {2a^2}   
\nonumber \\
\Gamma_x=\Gamma_y=0   \quad  .
\label{3dforces}
\end {eqnarray}
We now consider the   2D  case of  
mass  concentrated  on a half circle  of radius $a$ 
situated  on the positive  $y$-axis,
where now  
the mass per unit length $\mu_{2D}$ is
\begin{equation}
\mu_{2D} =  \frac{M}{\pi a}
\quad .
\end{equation}  
The  2D vectorial intensity  of the  field  is 
\begin{equation}
d {\bf \Gamma}  
=G \frac{dm}{a^2}
\quad  ,
\end{equation}  
being 
\begin{equation}
dm = \mu_{2D}  a   d \theta 
\quad ,
\end{equation}
where $0 \leq \theta \leq \pi $.
The 2D Cartesian gravitational components 
of the force at the center 
($x$=0,$y$=0) are  
\begin{eqnarray}
\Gamma_y =  G \frac{\mu_{2D}}{a} \int_0^{\pi} \sin \theta 
d \theta
     \nonumber \\
\Gamma_x =  G \frac{\mu_{2D}}{a} \int_0^{\pi} \cos \theta 
d \theta
\quad  .
\end {eqnarray}
The integration   of the 2D case   gives
\begin{eqnarray}
\Gamma_y = \frac{2 G M} {\pi a^2}   \nonumber \\
\Gamma_x=0   \quad  .
\label{forcecircle}  
\end {eqnarray}
At the moment of  writing  
the Committee on Data for Science and Technology (CODATA)
recommends 
\begin{equation}
G=(6.67384 \pm 0.00080)\times 10^{-11}
\frac { m^3} {kg
s^2}
\quad ,
\end{equation}
see \cite{Mohr2008}.
Before to continue we express the Newtonian 
gravitational constant  in the following units:
length  in $Mpc$,  mass in $M_{gal}$ which is
$10^{11} M_{\sun}$  and  $yr8$
which are $10^8$ $yr$
\begin{equation}
G=4.49975  10^{-6}
\frac { Mpc^3} {M_{gal}yr8^2}
\quad .
\end{equation}
The two formulas (\ref{3dforces}) in 2D  and 
(\ref{forcecircle})  in 3D represent a useful
reference  to test a numerical code 
and to fix the range of variability 
of the gravitational field.
According to our 3D theory  the  gravitational 
field at the center of the cosmic voids 
varies between the  minimum value of  zero (shell theorem)
and  a maximum value 
\begin{equation}
\Gamma_z = \frac{G M} {2{\overline{R}}^2}
=6.769\,10^{-9} N \, \,  
\frac { Mpc  M_{gal}}{yr8^2}
\quad , 
\end{equation}
where $N$  is the number of  galaxies 
in the spherical shell surrounding  the cosmic void
having  mass 
$M=M_{gal}$    and
${\overline{R}}$=18.23  Mpc is the average  radius  of 
the cosmic voids.

We are now ready  to process 
the 2MRS  data and we associate  to each  galaxy,
as reported in 
Figure \ref{2mrs_cut} a  mass given
by Eqn.  \ref{ratior}. 
The three  components 
of the gravitational  field  
are reported 
in  table  \ref{dataforces2mrs}.

\begin{figure}
\begin{center}
\includegraphics[width=6cm]{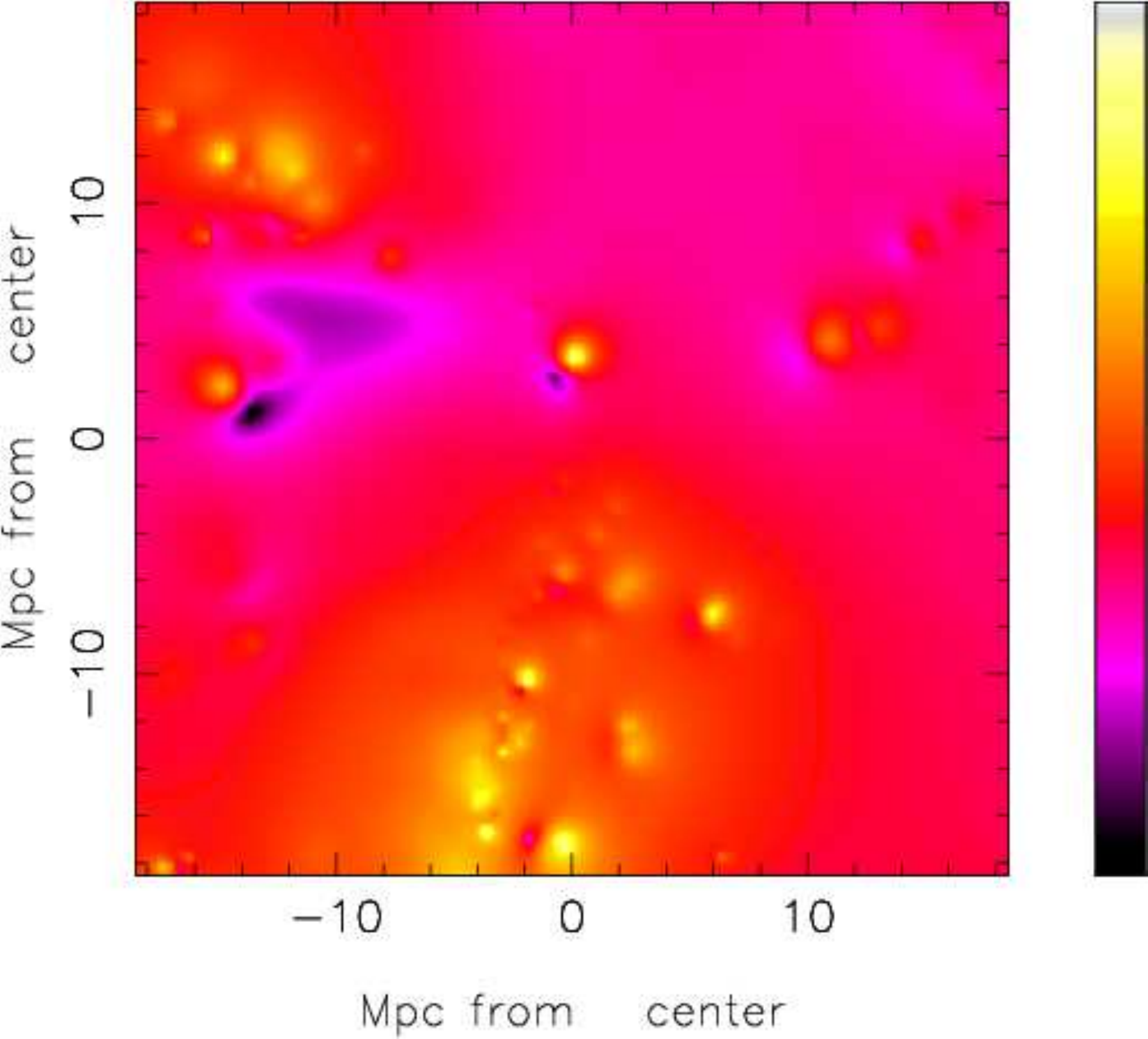}
\end {center}
\caption
{
Cut-map of the 3D gravitational field of  2MRS 
when  $R$=6. 
In order to  have periodic boundary conditions 
the side of the box  is 37   Mpc.
}
          \label{3dfield}%
    \end{figure}

\begin{table}
\caption { 
3D Gravitational forces
expressed in $\frac { Mpc  M_{gal}}{yr8^2}$
at the center of   a 3D box of side 
37 $\times$ 2 Mpc when  $R=$6
and theoretical 3D formula ( \ref{3dforces}).
At  z=0.008 the  efficiency of the sample is
$\approx$ 70.6 $\%$. 
}
 \label{dataforces2mrs}
 \[
 \begin{array}{ccccc}
 \hline
 \hline
 \noalign{\smallskip}
Environment   & \Gamma_x     & \Gamma_y   & \Gamma_z & \| \Gamma \| \\
 \noalign{\smallskip}
 \hline
 \noalign{\smallskip}
real~structure &-9.77\, 10^{-6} &-1.53\, 10^{-5} 
& -3.04 \, 10^{-6}  & 1.84 \,10^{-5}  \\  
half~sphere & 0 & 0 
& 2.55 \, 10^{-5}  & 2.55  \,10^{-5}  \\  
\noalign{\smallskip}
\noalign{\smallskip}
 \hline
 \hline
 \end{array}
 \]
 \end {table}
From a careful analysis of   Table~\ref{dataforces2mrs}
it is possible to conclude that the gravitational field 
is greater than zero  but smaller in respect to the case
in which all the galaxies resides on a half sphere of
radius equal to the averaged radius of the sample.
Figure  \ref{3dfield} reports a slice at the middle 
of a smaller box.

\begin{figure}
\begin{center}
\includegraphics[width=6cm]{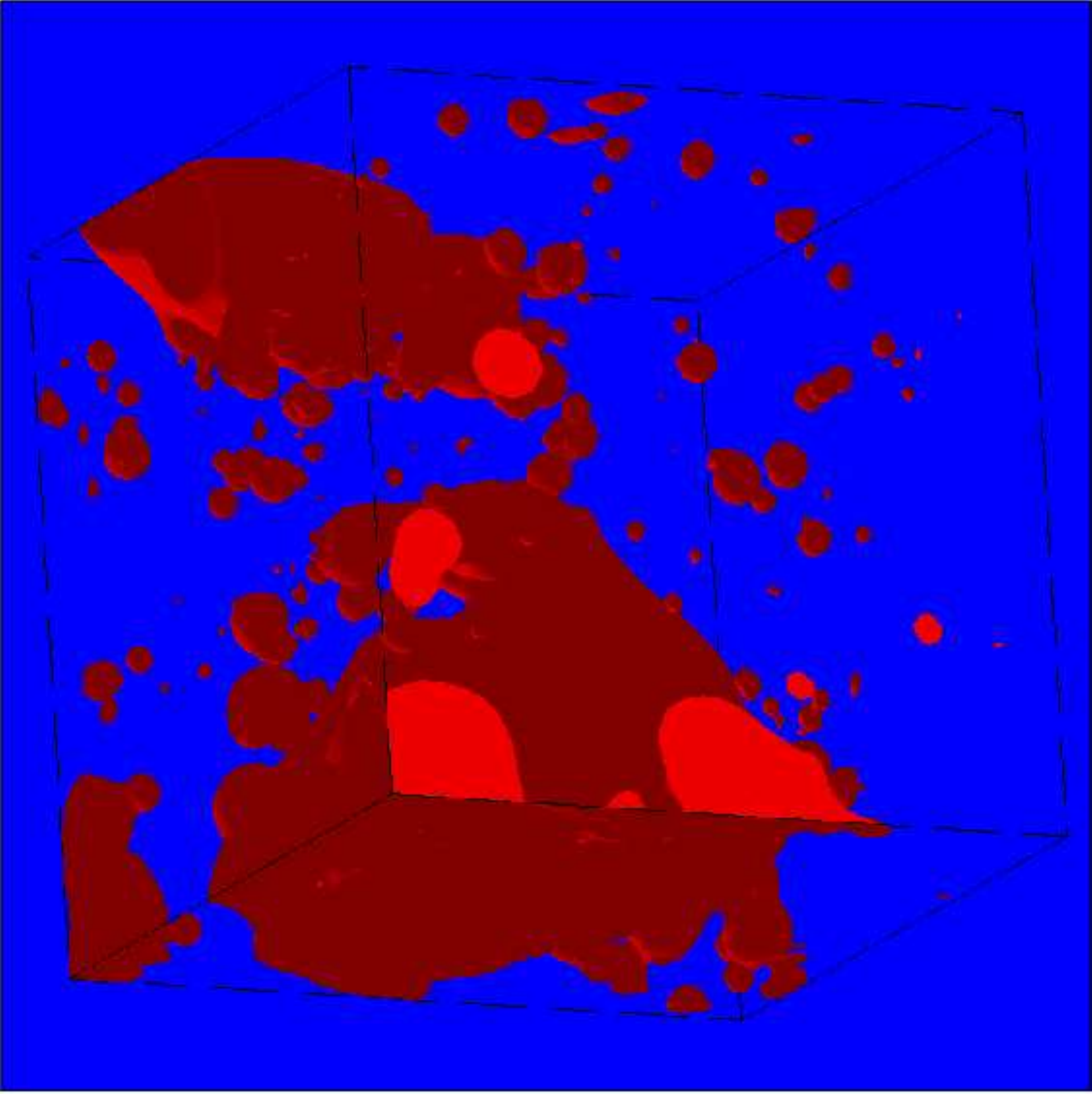}
\end {center}
\caption
{
Iso--surface of  the logarithm of the 3D 
gravitational field of  2MRS 
when  $R$=6 and $coef=0.43$. 
The orientation  of the figure is characterized 
by the Euler angles,
which are
     $ \Phi   $=30$^{\circ }$,
     $ \Theta $=30$^{\circ }$
and  $ \Psi   $=30$^{\circ }$.
}
          \label{iso3dfield}%
    \end{figure}

\begin{figure*}
\begin{center}
\includegraphics[width=10cm]{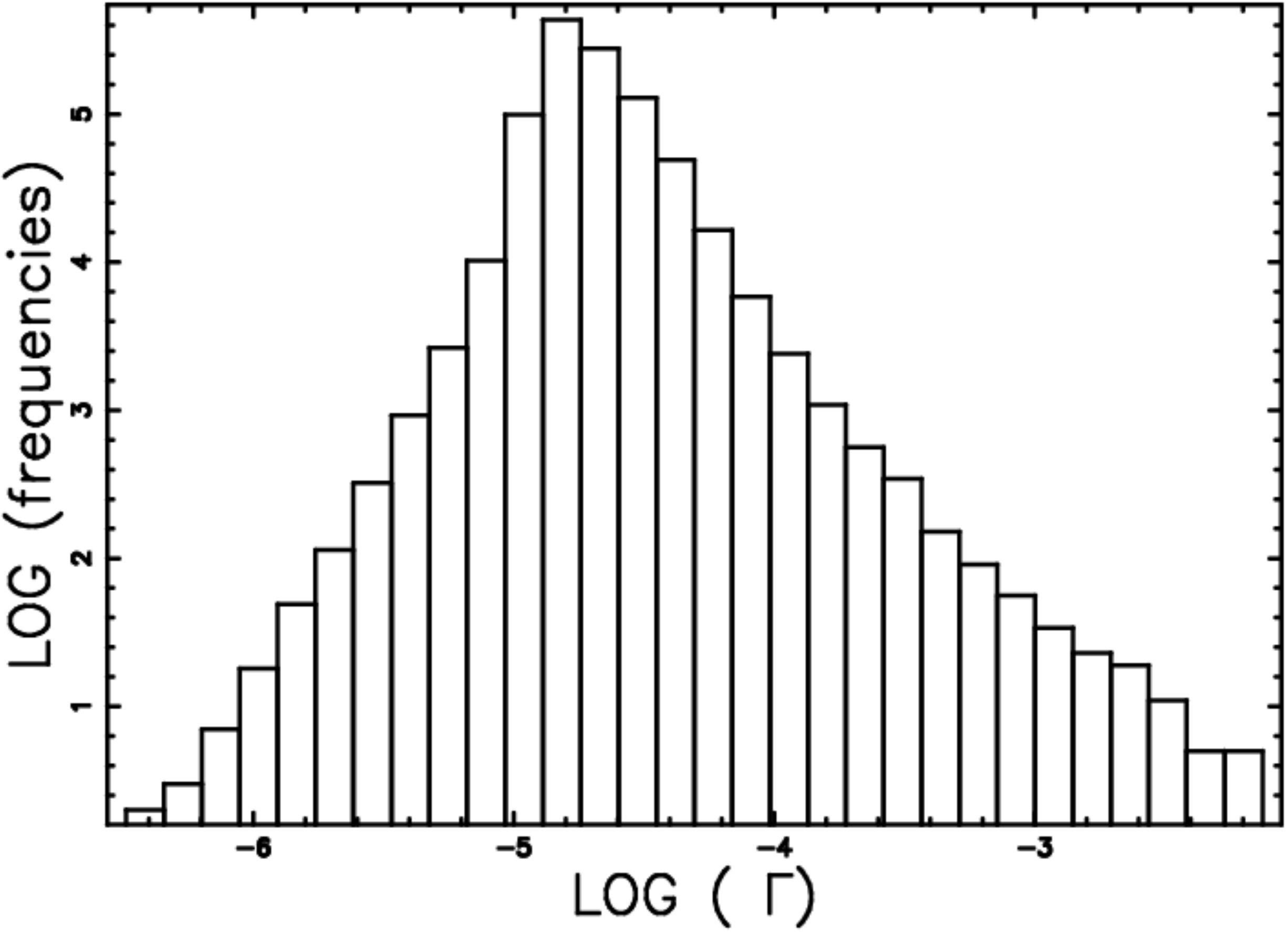}
\end {center}
\caption {Decimal  logarithmic histogram (step-diagram) 
of the values of the gravitational  field  
evaluated in  101$\times$101$\times$101 points.}
\label{gravisto}%
    \end{figure*}

The spatial displacement of the 3D grid
${ \Gamma }({i,j,k})$ 
which represent the absolute value of the 
gravitational field 
can be visualized
through the iso-density
contours, and as an example we considered
a 101$\times$101$\times$101 grid. 
In order   to do so,
the  maximum  value    ${ \Gamma_n }({i,j,k)}_{max}$
and the minimum  value ${ \Gamma_n }({i,j,k)}_{min}$~
should be extracted
from  the three-dimensional grid.
A value  of this grid can be   fixed by
      the following  equation:
\begin{equation}
\Gamma_n  (i,j,k)_{chosen} =
\Gamma_n  (i,j,k)_{min} +
(\Gamma_n (i,j,k)_{max} -
 \Gamma_n (i,j,k)_{min}) \times  {coef}
\quad,
\end{equation}
where {\it coef} is a parameter comprised between
0 and 1.
This iso-surface rendering 
of the gravitational field 
is reported in
Fig~\ref{iso3dfield}; the Euler
angles characterizing the point of view of the
observer are also reported.
Another  
interesting quantity to plot is the statistics 
of the values of the  already defined
spatial grid $\Gamma_n$    which holds 
101 $\times$ 101 $\times$101  values of gravitational field,
see Figure \ref{gravisto}.
From this histogram is possible to 
conclude that the 90$\%$ of the space 
has  a gravitational field  comprised 
$3.24~10^{-7} \frac { Mpc  M_{gal}}{yr8^2}  
\le \Gamma_n \le 3.16~10^{-5}
\frac { Mpc  M_{gal}}{yr8^2}  $.

\section{The Voronoi simulation}

\label{secvoronoi}
\begin{figure}
\begin{center}
\includegraphics[width=6cm]{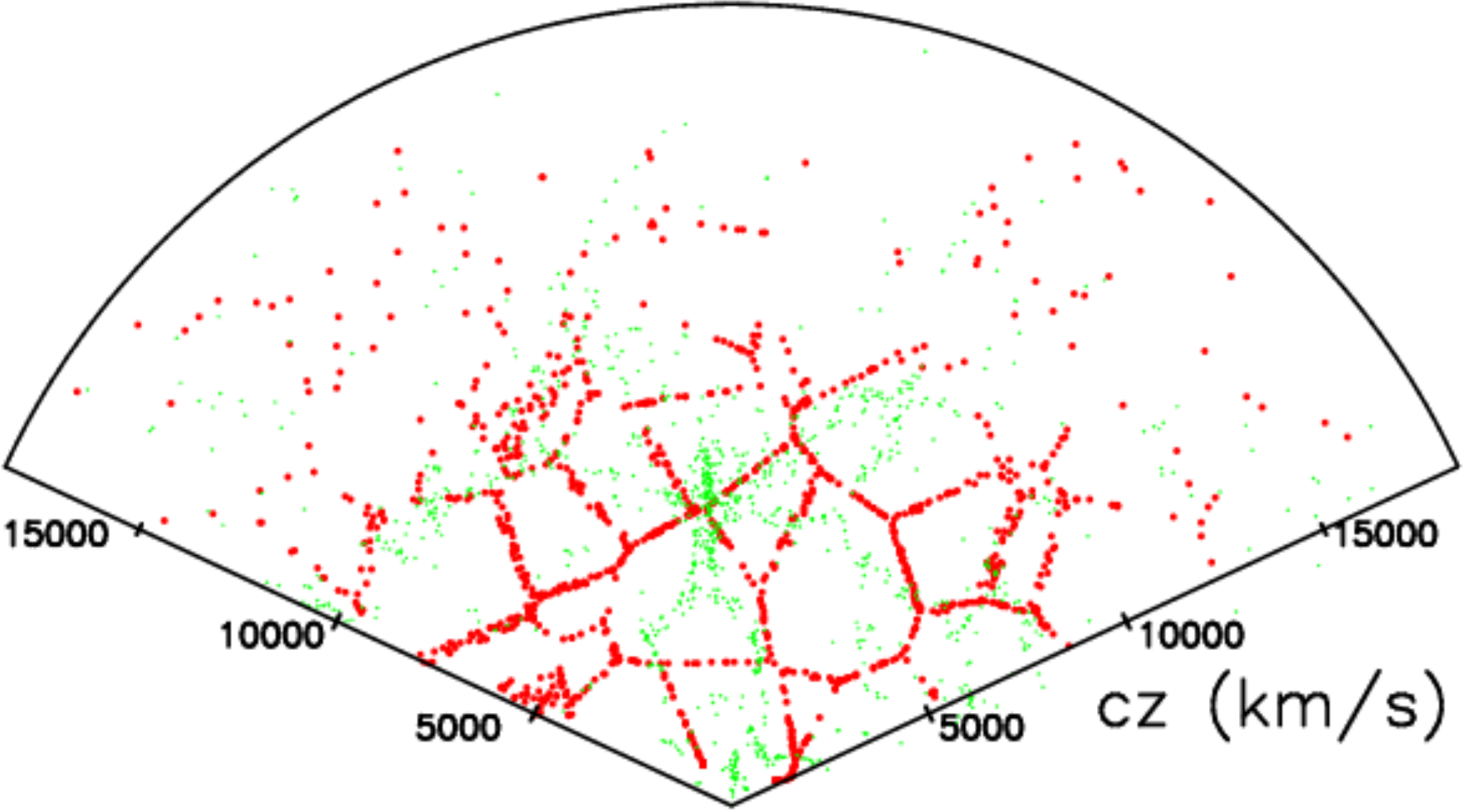}
\end {center}
\caption
{
Polar plot of   the   real galaxies (green  points)
belonging to the second CFA2 redshift catalog
and  the simulated galaxies in the  PVT 
framework (red points).
More details can be found  in 
\cite{Zaninetti2006}.
}
          \label{true_simu_color}%
    \end{figure}

The  Poisson Voronoi tessellation (PVT)  is 
a useful tool to explore the spatial 
clustering of galaxies.
The filaments  of galaxies visible 
in the slices-type   catalogs  are due  
to the intersection between a plane  and the PVT
network of faces as a first approximation.
An improvement can be obtained   by coding
the intersection between the slice of a given  opening 
angle  and the PVT network of faces, 
see \cite{Zaninetti2006,Zaninetti2010a}. 
As an example Figure \ref{true_simu_color}  
reports  both  the  CFA2 slice as  well the simulated slice.

\begin{figure*}
\begin{center}
\includegraphics[width=10cm]{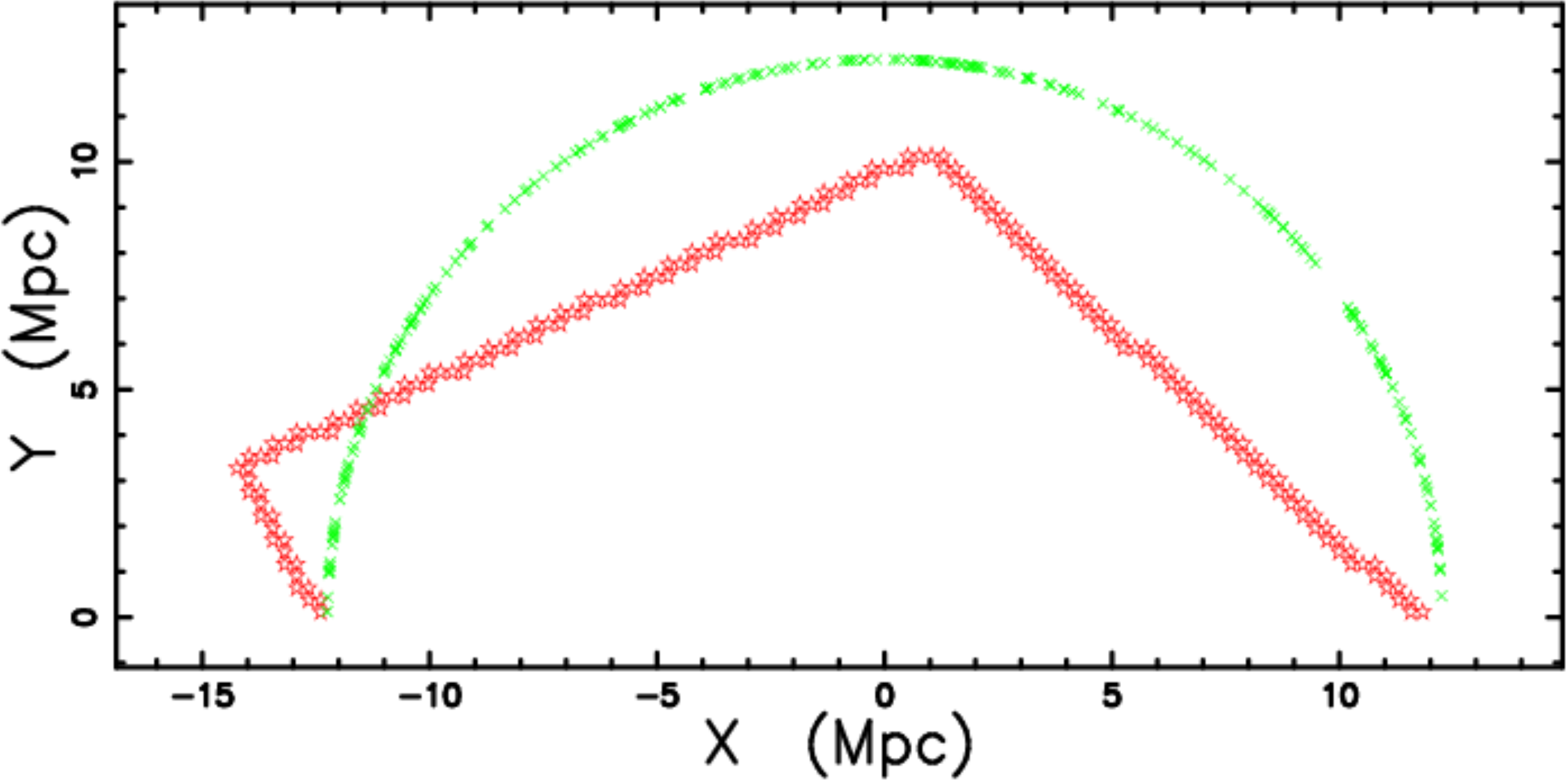}
\end {center}
\caption
{
Half circle  with $a$=12.25 Mpc     (green stars)  
and half irregular polygon (red squares).
The number of  unit  masses, 1 $M_{gal}$  is  455.
}
\label{semicerchio_due}%
    \end{figure*}

\begin{figure}\begin{center}
\includegraphics[width=10cm]{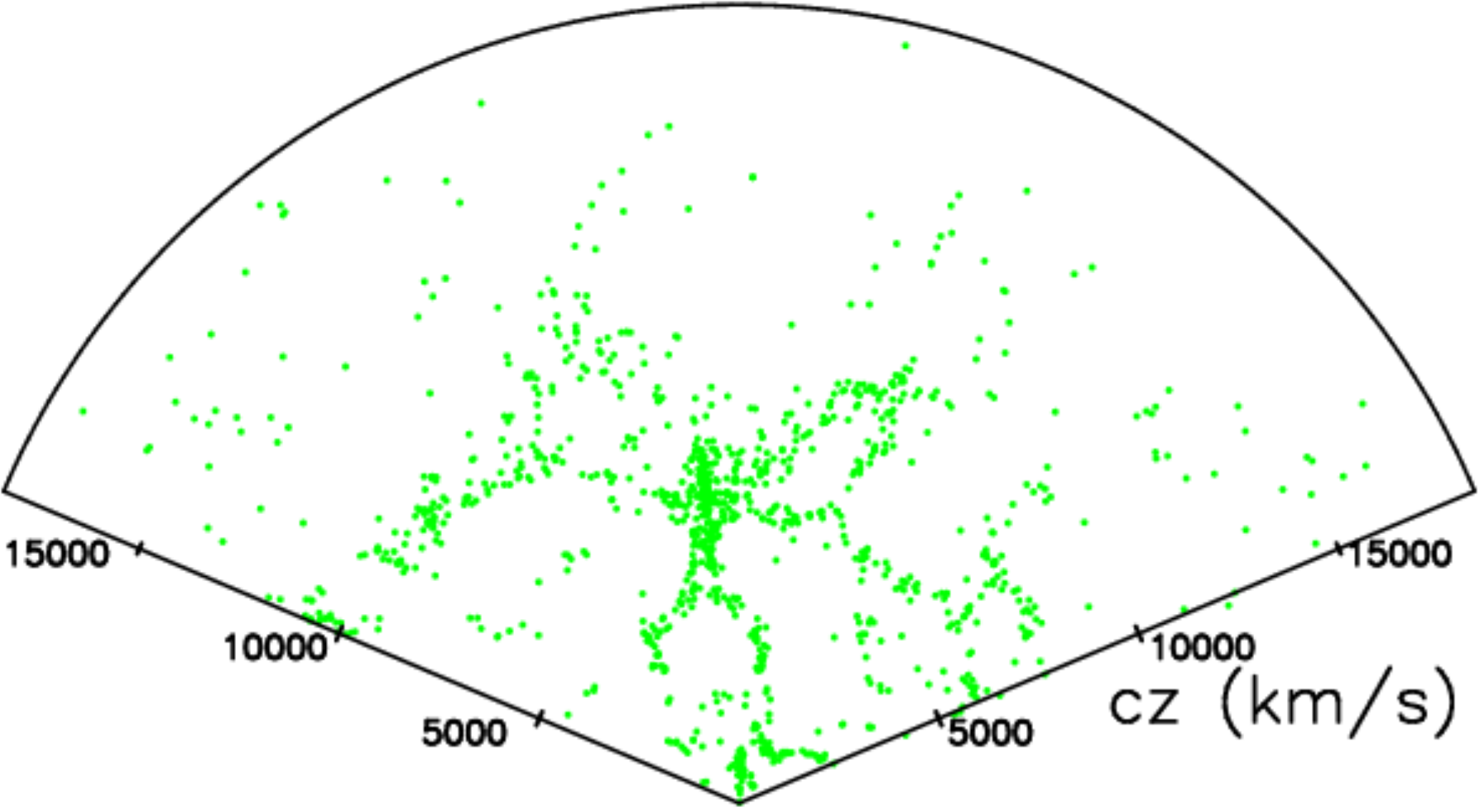}
\end{center}\caption 
{
Polar plot of   the   real galaxies (green  points)
belonging to the second CFA2 redshift catalog.
}
          \label{simu_cfa}%
    \end{figure}

We now test  formula  (\ref{forcecircle}) 
in a discrete environment rather than  in the  continuous case.
The test now  calculates the two forces $\Gamma_x$  
and $\Gamma_y$   in the center of the circle and in 
a 2D irregular Voronoi polygon generated  by PVT
which has the same averaged radius of the circle
and center occurring in the   same location of the 
generating seed.
The half  Voronoi polygon and the half circle
are  displayed in Figure \ref{semicerchio_due}
and the  two forces 
$\Gamma_x$  
and $\Gamma_y$  in Table \ref {dataforces}.
\begin{table}
\caption { 
Gravitational forces
expressed in $\frac { Mpc  M_{gal}}{yr8^2} $
in  the comparison between half circle, 
2D formula (\ref{forcecircle}),
and half irregular polygon.
The  parameters are $M_{gal}$=455 and $a=12.25453  Mpc$.    
}
 \label{dataforces}
 \[
 \begin{array}{ccc}
 \hline
 \hline
 \noalign{\smallskip}
Environment   & \Gamma_x     & \Gamma_y         \\
 \noalign{\smallskip}
 \hline
 \noalign{\smallskip}
Half ~circle -theory  & 0  & 8.67  \, 10^{-6}                \\  
Half ~circle -numeric & -4.16\, 10^{-8}  & 8.81 \, 10^{-6} \\  
Half ~polygon-numeric &1.37\, 10^{-6}  & 1.36 \, 10^{-5} \\  
\noalign{\smallskip}
\noalign{\smallskip}
 \hline
 \hline
 \end{array}
 \]
 \end {table}
We are now ready  to process a real void 
and our attention is focused  on a CFA2 slice
visible  in Figure \ref{simu_cfa}.

\begin{figure*}
\begin{center}
\includegraphics[width=10cm]{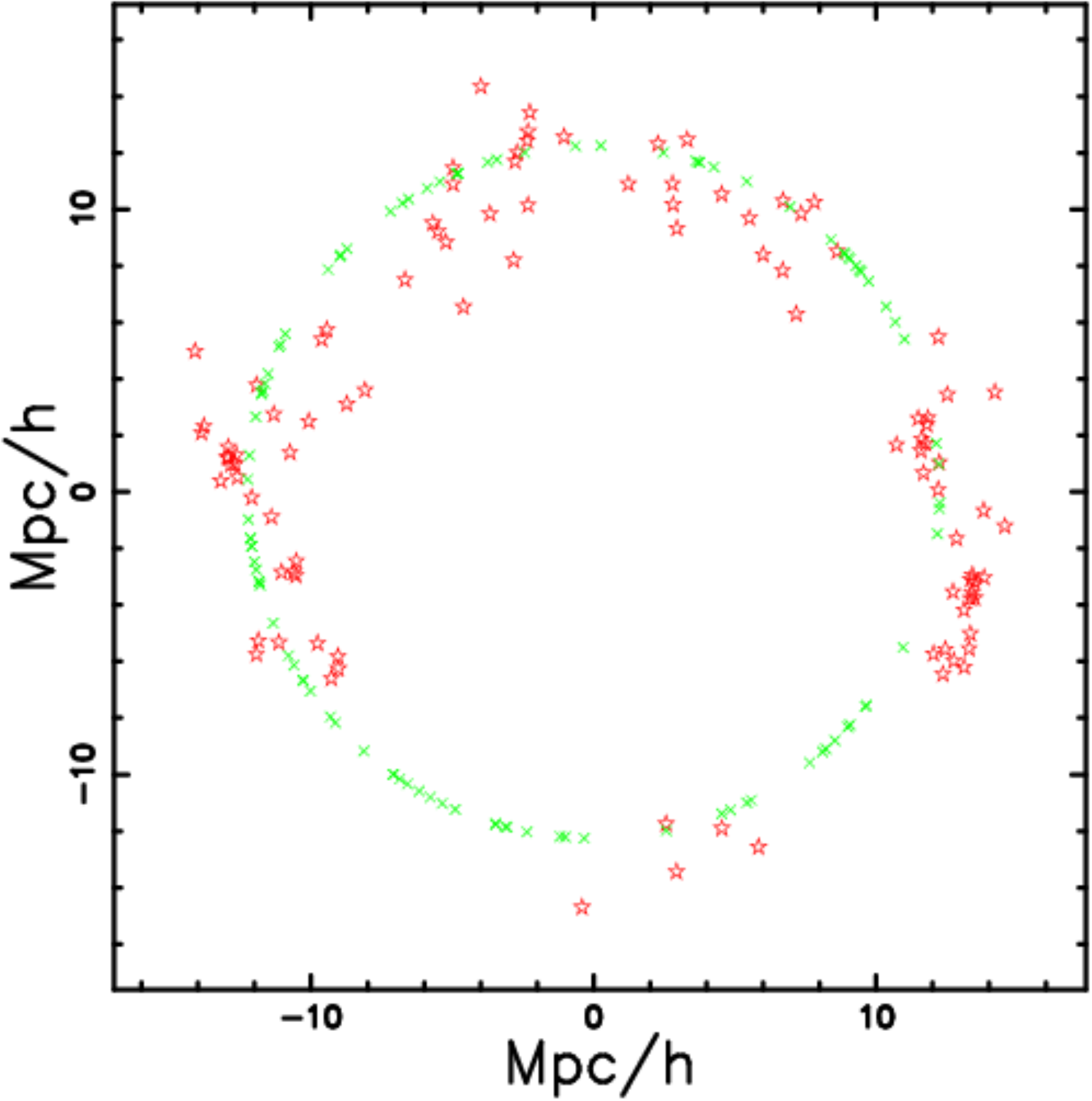}
\end {center}
\caption
{
Circle  with $a$= 12.25 Mpc    (green stars)  
and real void extracted from a CFA2 slice (red squares).
The number of  galaxies with  unit  mass  is  101.
}
\label{cerchio_vuoto}%
    \end{figure*}

A real void is extracted  and the averaged radius of the  
galaxies on the boundary of that void  is computed, see 
Figure \ref{cerchio_vuoto}.
The forces in the $x$ and $y$ direction are  then 
computed and  reported in Table~\ref{dataforcescfa2}.
\begin{table}
\caption { 
Gravitational forces
expressed in $ \frac { Mpc  M_{gal}}{yr8^2}  $
for the comparison between a   circle
and a real CFA2 void.
The  parameters are $M_{gal}$=101 and $a=12.25453  Mpc$.    
}
 \label{dataforcescfa2}
 \[
 \begin{array}{ccc}
 \hline
 \hline
 \noalign{\smallskip}
Environment   & \Gamma_x     & \Gamma_y         \\
 \noalign{\smallskip}
 \hline
 \noalign{\smallskip}
circle--theory           & 0  & 0                \\  
half~circle--theory      & 0  & 1.92\,10^{-6}    \\  
real~void--numeric       & -1.36\, 10^{-7}  & 8.3 \, 10^{-7} \\  
\noalign{\smallskip}
\noalign{\smallskip}
 \hline
 \hline
 \end{array}
 \]
 \end {table}

The presence  of both  a discrete  number of galaxies  
and  a non exactly
symmetric displacement of 
the galaxies produces   gravitational forces that take 
a finite value rather than zero.
It is interesting to point out that $\Gamma_y$  due to 
the galaxies on the boundary of the
real void   is  smaller than the theoretical  value
as given by the half circle which  represents
a  maximum theoretical value.

\section{Conclusions}

The  masses  of the galaxies can be deduced starting from 
the luminosities in the framework  of the mass 
luminosity ratio, $R$.
The spatial  distribution of the masses of the galaxies 
allows the computation  of the Newtonian gravitational
forces on the unit mass.
As a reference for the evaluation  of the forces the 
2D and 3D shell theorem is analyzed.
The evaluation of forces at the center 
of the box allows to conclude that the forces are smaller 
in respect to the mass concentrated on  a half sphere
of radius equal to the averaged radius 
of the selected sample of galaxies
but bigger than zero due to the fact that the distribution 
of the galaxies is discrete rather than continuum.
A careful analysis   of a cubic box 
having  side of 37 Mpc allows to say that the 90 $\%$ of
the space has gravitational forces
around the average value of 
$ 2.1\,10^{-5} \frac { Mpc  M_{gal}}{yr8^2}$.


\end{document}